
\magnification=1200

\def\.{\mathaccent 95}
\def\a{\alpha}
\def\be{\beta}
\def\ga{\gamma}

\def\ka{\kappa}

\def\si{\sigma}

\def\Ga{\Gamma}

\def\frac#1#2{{\textstyle{{#1}\over {#2}}}}

\def\lsim{\mathrel{\rlap{\lower4pt\hbox{\hskip1pt$\sim$}}
    \raise1pt\hbox{$<$}}}
\def\gsim{\mathrel{\rlap{\lower4pt\hbox{\hskip1pt$\sim$}}
    \raise1pt\hbox{$>$}}}
\def\sqr#1#2{{\vcenter{\vbox{\hrule height.#2pt
         \hbox{\vrule width.#2pt height#1pt \kern#1pt
         \vrule width.#2pt}
         \hrule height.#2pt}}}}

\newbox\grsign \setbox\grsign=\hbox{$>$} \newdimen\grdimen \grdimen=\ht\grsign
\newbox\simlessbox \newbox\simgreatbox
\setbox\simgreatbox=\hbox{\raise.5ex\hbox{$>$}\llap
     {\lower.5ex\hbox{$\sim$}}}\ht1=\grdimen\dp1=0pt
\setbox\simlessbox=\hbox{\raise.5ex\hbox{$<$}\llap
     {\lower.5ex\hbox{$\sim$}}}\ht2=\grdimen\dp2=0pt

%
%

\def\ref#1  {\noindent \hangindent=24.0pt \hangafter=1 {#1} \par}
\def\doublespace {\smallskipamount=6pt plus2pt minus2pt
                  \medskipamount=12pt plus4pt minus4pt
                  \bigskipamount=24pt plus8pt minus8pt
                  \normalbaselineskip=24pt plus0pt minus0pt
                  \normallineskip=2pt
                  \normallineskiplimit=0pt
                  \jot=6pt
                  {\def\smallskip {\vskip\smallskipamount}}
                  {\def\medskip   {\vskip\medskipamount}}
                  {\def\bigskip   {\vskip\bigskipamount}}
                  {\setbox\strutbox=\hbox{\vrule
                    height17.0pt depth7.0pt width 0pt}}
                  \parskip 12.0pt
                  \normalbaselines}

\def\ts{\times}
\def\lb{\langle}
\def\rb{\rangle}

$${\bf RECONNECTING\ MAGNETIC\ FLUX\ TUBES\ AS\ A\ SOURCE\ OF}$$
$${\it IN\ SITU}{\bf \ ACCELERATION\ IN\ EXTRAGALACTIC\ RADIO\ SOURCES}$$

\centerline{Eric G. Blackman}
\centerline{Institute of Astronomy, Madingley Road, Cambridge CB3 OHA, England}
\medskip
\centerline{accepted for publication in {\it Astrophysical Journal Letters}}

\medskip
\doublespace

$$\bf ABSTRACT$$

Many extended extragalactic radio sources require a local $in\ situ$
acceleration mechanism for electrons, in part because the
synchrotron lifetimes are shorter than the bulk travel time across the
emitting regions.  If the magnetic field in these sources is
localized in flux tubes, reconnection may occur between regions
of plasma $\be$ (ratio of particle to magnetic pressure) $<<1$,
even though $\beta$ averaged over the plasma volume may be $\gsim 1$.
Reconnection in low $\beta$ regions is most favorable
to acceleration from reconnection shocks. The reconnection X-point
regions may provide the injection electrons for their subsequent
non-thermal shock
acceleration to distributions reasonably consistent with observed
spectra.  Flux tube reconnection might therefore be able to provide
$in\ situ$ acceleration required by large scale jets and lobes.
\bigskip
\noindent ${\bf Subject\ Headings:}$ Magnetic Fields: MHD;
Acceleration of Particles; Galaxies: Active, Jets, Magnetic Fields

\vfill
\eject

$\bf I.\ Introduction$

The non-thermal emission from extended extragalactic radio sources is
likely synchrotron radiation by
relativistic electrons (Begleman et al., 1984).
Observationally, the
flux satisfies $F{(\nu)}\propto \nu^{-\alpha},$
where $\a$ is the spectral index, and $\nu$ is the frequency.
This corresponds to a power law electron distribution with a number density
$N(\ga_e)\propto \ga_e^{-s},$
where $\ga_e$ is the electron Lorentz factor and $s=2\a+1$.
For over 90\% of the sources, $0.5 < \alpha < 1$ (Begelman et al., 1984),
corresponding to  $2 < s  <3$.
Jets and hot spots show flatter spectra than  diffuse radio lobes.

Many large scale synchrotron sources have sizes larger than
the distance electrons can be convected in a synchrotron loss time,
exhibiting the need for an $in\ situ$ acceleration
process (Achterberg, 1987).
When the magnetic field is sufficiently tangled, electron transport
is due to the bulk flow.
The transport distance in a synchrotron loss time is
$$D_{syn}\equiv V_b\tau_{syn}=V_b\ga_e {m_e}c^2/P_{syn}=(8 \times
10^8)V_b /\ga_e B^2,\eqno(1)$$
where $P_{syn}$ is the synchrotron power, $m_e$ is the electron mass,
$B$ is the magnetic field magnitude,  and $V_b$ is the bulk flow
speed.
An electron of energy $m_ec^2\ga_{e}$
radiates according to (Rybicki \& Lightman, 1979)
$$\nu=1.6\times 10^{-2} (B/10^{-6}{\rm G})(m_ec^2\ga_{e}/{\rm GeV})^2\ {\rm
GHz}.\eqno(2)$$
Using $(1)$ and $(2)$, an electron in a Mpc scale lobe radiating at $1.5$ GHz
in a $5$ $\mu$G field has
$\ga_e\sim 7.7\times10^3$,  and
will lose its energy in $>$ 10 Mpc if
$V_b =c$.  However, if a thermal plasma is present in the lobes,
plasma instabilities can reduce the bulk velocity to  $V_b<c$
(Achterberg, 1987).
For $\beta\equiv P/ P_B\lsim 1 $, where $P$ is the particle
pressure and $P_B$ is the magnetic field pressure,
the maximum becomes the Alfv\'en speed
$V_{A}\sim 5\ts 10^7 (B/{5\times 10^{-6} {\rm G}})(n/10^{-4}{\rm
cm^{-4}})^{-1/2}\ {\rm cm\ s^{-1}},$
where $n$ is the number density.
Thus, $D_{syn}\lsim 20$ kpc. Even more suggestive
are observations of optical synchrotron jet emission, for example in
3C33 (Meisenheimer and R$\ddot {\rm o}$ser, 1986), which imply
$\sim$ 100 GeV electrons ($\ga\sim
10^5$)  radiating over several kpc
in an equipartition (EQP) field of 10$^{-3}$G.
The synchrotron lifetime is then $\sim 200$yr, much shorter than
the jet length travel time, so that $in\ situ$
acceleration is required for the jet emission to last.
HST observations support this (Macchetto, 1992).
In addition, many jets do not show the spectral line index gradients
expected from expansion or synchrotron losses (Achterberg, 1987).

The energy for particle acceleration and turbulence
is converted from the bulk flow.
Because the turbulent
motions are responsible for stretching the magnetic field, the
average magnetic field energy will be at most equal to the average
turbulent energy.
These points are summarized for primarily non-relativistic bulk flows by
$$(1/2)\rho_{ave} V_{b}^2 \ge (1/2)\rho_{ave} V_L^2 \ge  \lb B^2\rb/8\pi\ \
{\rm
and}\ \ (1/2)\rho_{ave} V_{b}^2 \ge P\sim \rho_{e} c^2\lb\ga_e\rb,\eqno(3)$$
 where $\rho_{ave}$ and $\rho_e$ are the  average and electron
 densities, $V_L$ is the outer eddy
scale velocity and the brackets indicate the average.
The presence of a large scale mean flow and mean magnetic
field does not preclude
the presence of a tangled small scale field.  In the Galaxy for
example, the rotation provides an underlying ordered flow,
while the observed large and small scale fields have the same
magnitude (Heiles, 1995).

Previous studies have appealed to stochastic and
fast shock acceleration (Achterberg, 1987; Eilek \& Hughes, 1993).
The first results directly from turbulent motions in
magnetized plasmas or from low frequency plasma and MHD waves.  The second
occurs near fast shocks that may be present in the flow.
It is possible for both mechanisms to accelerate particles to power law
spectra.  However, both are ``re-acceleration'' mechanisms;
electrons need to be pre-accelerated before these processes can
maintain high energy electron tails.

In this paper, a kind of hybrid approach is  suggested
which appeals to $slow$ shocks that are naturally produced
from magnetic reconnection events (for an alternative approach
to reconnection in jets, see Romanova \& Lovelace, 1992).
The shocks would occur throughout the plasma, wherever a reconnection
site occurs.  Reconnection shocks also have the unique feature
of an adjacent X-point region where the field annihilates.  This
region may be able to directly inject electrons to their required
pre-accelerated energies.
Section II addresses how reconnection
shocks can accelerate electrons.  The most efficient shock
acceleration, with power law indices appropriate for large
scale jets, would occur when the reconnection takes place in regions of low
plasma $\beta$ (Blackman \& Field, 1994).
In section III, I point out that this may occur in an astrophysical
plasma even if the average ratio of the particle to magnetic pressure
$\beta_{ave}\gsim 1$, when the reconnection occurs between reduced
density flux tubes.  In Section IV, I address the application to
jets and lobes more specifically, and summarize in section V.

$\bf II.\ Reconnection\ Shock\ Acceleration$

Magnetic reconnection occurs
as regions of skewed magnetic polarity intersect.
The intersection  produces a
thin dissipation region  where flux freezing is
violated and magnetic field is annihilated (e.g. Biskamp, 1994).
The annihilation produces a topology change with an
``X-point'' at the interface.  Much of the work in reconnection
theory has focused on determining how fast oppositely magnetized plasma flows
can merge across an X-point region in the steady-state.
Although the dependence of this rate on $R_M$ has been
debated, many simulations show the presence of slow mode
shocks (Biskamp, 1994), highlighting the abrupt change in direction
between the inflow to the X-point and outflow from the X-point.
Such slow shocks are distinguished from fast mode shocks in part by the
fact that the magnetic field downstream of a slow shock has a lower magnitude
than its value upstream---the opposite to the fast shock.

Blackman \& Field (1994), show that low $\be$
reconnection slow shocks are the strongest slow shocks, giving
compression ratios $q$ of
$2.5<q\equiv\rho_d/\rho_u<4,$
where $\rho_{d(u)}$ is the
downstream (upstream) density (see also Kantrowitz \& Petschek, 1966).
The upper limit corresponds to a quasi-parallel
shock (upstream magnetic field  nearly parallel to the shock normal),
and the lower limit corresponds to a quasi-perpendicular shock
(upstream magnetic field  nearly perpendicular to the shock normal).
The importance of $q$ can be seen from
the analytical approach to shock acceleration.
 Assuming  that gradients in the
normal direction $>>$ those along the shock,
the diffusion-convection equation across a shock is given by (Jones \& Ellison,
1991)
$$\partial_n[V_n N-\ka_n \partial_n N]-
(1/3)(\partial_n V_n)\partial_{p_p}[p_pN]=0,\eqno(4)$$
where $V_n$ is the normal flow velocity across the shock, $\ka_n$ is the normal
diffusion coefficient, and $p_p$ is the particle momentum.
Fermi acceleration operates as the particles diffuse between
scattering centers (presumably Alfv\'en turbulence) on each side of the shock.
Particles always see the centers converging, as the normal velocity is
larger upstream.  The solution of $(4)$
across the shock with thickness $<<$ mean free path (Jones \& Ellison, 1991)
indeed
shows that the outflow energy spectrum for a steeper inflow  spectrum takes the
power law form $N\propto \ga_e^{-s}$, with energy index
$s=(q+2)/(q-1)$.
For $2.5<q<4$ we then have $2<s<3$ for these shocks, which is
consistent with the observed range for extragalactic radio sources
given in section I.

Though an analytical treatment was employed above,
shock acceleration is actually a very non-linear process; the
Fermi acceleration engine across shocks is very efficient,
transferring $\ge 1/5$ of the inflow energy to particles (Jones \& Ellison,
1991).
These particles tend to smooth out the
shock by diffusion, produce turbulence,
and can even increase the compression ratio above the
jump condition value, as their escape and acceleration
change the downstream equation of state.
Non-thermal acceleration is $enhanced$ in the non-linear regime.
For an ion-electron plasma,
the signature of effective non-linear shock-Fermi acceleration
in simulations is the appearance of
electromagnetic beam instabilities (Jones \& Ellison, 1991),
brought on by the interaction of
back-scattered, energized, ions with the inflowing plasma.
The instabilities indicate that
non-thermal ions are produced from an initially thermal input,
initiating the Fermi acceleration process.
Such instabilities have been seen in fast as well as
in slow shock simulations (Omidi \& Winske, 1994).


Some observational support for reconnection shock acceleration
is present in the geomagnetic tail where
turbulence, required for Fermi acceleration, is seen on both sides of the
shock fronts (Coroniti, et al., 1994).
In addition, although non-thermal ions are observed,
non-thermal tails in the electron spectra are also
seen (Feldman et al., 1990).  These
have not been modeled by hybrid simulations which assume a fluid
electron population.

It is indeed the acceleration of electrons, not ions, that is of interest
for radio sources.  Simulations show electron
acceleration, but only for electrons that are injected above some critical
energy (Ellison, 1992).  Slow shocks would operate similarly
with respect to Fermi acceleration.  (For relativistic flows,
Fermi acceleration would only operate for parallel shocks, while
for perpendicular shocks, non-thermal electron acceleration
requires a small proton fraction even for a primarily electron-positron jet
 [Hoshino et al., 1992]).
Injection is needed because Fermi acceleration demands that
the downstream particles be able to diffuse upstream.
Particles must have enough energy to resonantly
interact with the plasma waves that provide pitch-angle scattering.
For Alfv\'en turbulence (Eilek \& Hughes, 1991), the electron lower
bound is a factor
$\sim m_p/m_e$ times that for protons and is given by
$$\ga_e\gsim 1+(m_p/m_e)(V_{Ad}/c)^2\eqno(5)$$
for $ V^2_{Ad} < (m_e/m_p)^2 c^2$ while for $V^2_{Ad}>(m_e/m_p)^2c^2$,
$$\ga_e\gsim (m_p/m_e)(V_{Ad}/c),\eqno(6)$$
where $m_p$ is the proton mass, and
$V_{Ad}$ is the downstream Alfv\'en speed.

For reconnection shocks, the associated highly dissipative energy
conversion near the X-point
may provide the injection electrons.  To see this,
note that upon absorbing annihilated field energy at the X-point,
the average $\gamma_e$ there
$\sim 1+(V_{Au}^2/c^2)(m_p/m_e)$, where
$V_{Au}$ is the upstream Alfv\'en speed. (Note that
when $B_{Au}^2/4\pi \rho >c$, then $V_{Au}$
is replaced by  $\ga_{Au}V_{Au}$ where $\ga_{Au}$ is the associated Lorentz
factor.)  Since $\ga_{Au}V_{Au}>\ga_{Ad}V_{Ad}$ for a slow shock
(Blackman \& Field, 1994),
we see that an X-point region
can therefore in principle $always$ inject for both cases $(5)$ and $(6)$.



To solve  the $in\ situ$ acceleration problem for radio
sources, shock acceleration must
dominate synchrotron loss.  Thus,
$\tau_{syn}$ must  exceed the shock acceleration time $\tau_{sh}$:
$$\tau_{syn}=\ga_em_ec^2/P_{syn}= 6\pi m_ec/(\ga_e B_u^2 \si)>\tau_{sh}\sim
\kappa_{n}/V_u^2,\eqno(7)$$
where $V_u$ is the upstream velocity,
$\si$ is the Thomson cross section,
and $\kappa_{n}$ is the diffusion coefficient normal
to the slow shock.  For electrons moving at $c$, $\ka_{n}\sim
c(\ga_e m_e c^2/e B_d)$ (Achterberg, 1987).
{}From $(7)$, the condition
for $\tau_{syn}> \tau_{sh}$
(also justifying the absence of a synchrotron loss term in $(4)$)
is then
$$\gamma_e\lsim 3.8\times 10^8(V_u/10^8{\rm\ cm/s})(B_u/10^{-6}{\rm
Gauss})^{-1/2}{\rm Cos} \theta \eqno(8)$$
where $B_d \sim  B_u{\rm Cos}{\theta}$ (Blackman \& Field, 1994),
has been used, and $\theta$ is the
angle of the upstream field with respect to the shock normal.

$\bf III.\ Role\ of\ Flux\ Tubes\ for\ \be_{ave}\gsim 1$

Shock acceleration as described above, may occur
between regions of low $\beta$ plasma even when $\beta_{ave}\gsim 1$,
if the field is confined to flux tubes.
Vishniac (1995) considers the steady-state
dynamics of flux tubes in a turbulent
plasma with $(1/2)\rho V_L^2\sim\lb B^2\rb/8\pi$ and $\beta_{ave}>>1$.
In this case, the flux tubes fill a negligible fraction of the total volume.
However, if the relations in $(3)$ are near equalities, then
$\beta_{ave}\sim 1$. Here the flux
tubes would fill 1/2 the total volume.
As this regime is outside of the scope of Vishniac (1995),
it will be considered in more detail below.  It must be noted that
in neither Vishniac 1995,
nor the present paper, is the formation of flux tubes
solved as an initial value problem.
The presence of a steady-state is assumed and the consequences are explored.

For jets, Kelvin-Helmholtz instabilities between the jet flow and the
ambient medium (eg. Birkinshaw, 1991) are a source of turbulent eddies.
Each energy containing (outer) scale eddy of
wavelength $L$ would stretch a tube to length $L$ and radius $r_t$, and
the steady-state plasma would then contain a mesh of reconnecting
flux tubes.
The thickness of each tube, $r_t$, can be estimated by
balancing  the magnetic and  turbulent eddy
drag forces (to be justified later)
 (Landau \& Lifshitz, 1987;
Vishniac, 1995)${^1}$.  This gives
$$(B_t^2/4\pi r_c)(\pi r_t^2)\sim C_d\rho_{ext} V_L^2 2r_t,\eqno(9)$$
where $B_t$ is the magnitude of the field in the flux tube,
$\rho_{ext}$ is the density outside the
flux tube, and  $C_d$ is the coefficient of turbulent drag.
Since $L$ is a wavelength, the radius of curvature $r_c$ can be estimated
by $L/4$ when the tube maximally responds to the turbulence.  In equilibrium,
$B_t^2/8\pi\sim P_{ext}$, where $P_{ext}$ is the
external pressure, so that $(9)$ gives
$$r_t=L C_d \Ga M_L^2 /4\pi \sim L C_d/2\pi
\sim  0.4L/2\pi\sim L/16, \eqno(10)$$
where $M_L^2\equiv V_L^2/(\Ga P_{ext}/\rho_{ext}) \sim M_L^2 \sim 2/\Ga$, in
EQP
and $\Ga$ is the adiabatic index.
For the last similarity in $(10)$, $C_d$
was estimated from the ``drag'' crisis (Landau \& Lifshitz, 1987) which
reduces $C_d<1$ at large $R_L$.  Assuming $R_L\gsim 1000$, acceptable
for jets and lobes (Begelman et al., 1984), $C_d\sim 0.4$.

The steady-state MHD solution for the interior of a flux tube with
large aspect ratio (Vishniac, 1995)
shows that the density in a tube falls
off rapidly from the outside within a thin shell of thickness
$r_s\sim (\nu_M \tau_{ed})^{1/2}\sim L/R_M^{1/2}$,
where  $\tau_{ed}\equiv L/V_L$,
$\nu_M$ is the magnetic diffusivity, and $R_M$ is the magnetic
Reynolds number.  Inside the skin layer,
the Alfv\'en speed increases rapidly toward the center,
due to the reduction of the tube central density.
The tube reconnection time scale, $\tau_{rec}$, is the time to
reconnect the skin layer plus the time to reconnect half
the internal section of a tube.
Since $r_s<<r_t$ we have
$$\tau_{rec}=F(R_M)(r_s/V_{As}+r_t/V_{Ai})
\sim F(R_M) [L/(R_M^{1/2}V_{As})+L/(16 V_{Ai})],\eqno(11)$$
where $V_{Ai}=V_{Au}$ is Alfv\'en speed inside the tube,
$V_{As}\sim V_L$ is the Alfv\'en speed within $r_s$, and
$F(R_M)$ is either $R_M^{1/2}$ for Sweet-Parker reconnection, or
Log$R_m$ for Petschek reconnection (Parker, 1979).  We need to estimate
$V_{Ai}$, and thus the mass fraction in the tubes,
to obtain $\tau_{rec}$ and $V_{rec}=V_u=r_t/\tau_{rec}$.


To estimate the tube
reconnection velocity, and show that $\beta_t<<1$, justifying the use of the
{\it strong} compression ratios as given in section
II, it is first necessary to  distinguish between the $\be_{ave}>>1$
case of Vishniac (1995)
with the $\be_{ave}\sim 1$ case likely relevant for jets.
In the former, tubes are nearly completely
evacuated, and far from their nearest neighbor
while for the latter, each flux tube would be in contact with $\sim L/2r_t \sim
8$ other tubes because approximately 1/2 the volume is
filled with flux tubes.  The magnetic field structure
would therefore show a close mesh of reconnection sites.
Each reconnection site occurs where a tube flattens up
against a partner.  The dissipation region at each site
has a volume of $W_{rr} \sim (2 r_s)(2 r_t)^2$.
Because  magnetic field is annihilated in each dissipation region,
some material from the outside can leak in and slip
onto tube field lines there; every field line in the tubes must eventually
pass through X-point regions.  A full reconnection event requires a
height of $2r_t$ to reconnect, so the amount of matter that
loads into a given tube during each $\tau_{rec}$
is $\sim N_{rr} \rho_{ext}W_{rr}(2r_t/2r_s)$,
where $N_{rr}\sim 8$ is the number of reconnection sites per tube.
Since reconnection of closed
loops removes 1/2 of the matter in any tube during $\tau_{rec}$,
only 1/2 of the loaded matter remains.
The steady-state mass fraction of matter in the flux tubes is then
$(1/2)N_{rr} \rho_{ext}W_{rr}(2r_t/2r_s)/[\rho_{ext}L^3]
\sim 32 r_t^3/L^3\sim 4\times 10^{-3},$
using $(10)$.
Since the tubes comprise 1/2 the total volume, the density in the tubes is
then $\rho_t \sim 4\times 10^{-3}\rho_{ext}$, and though the
tubes are not completely
evacuated, $\be_t\sim 4\times 10^{-3}\be_{ave}<<1 $,
where $\beta_t$ is $\beta$ inside the tubes.
Given the density reduction in the tubes,
 $V_{Ai}=V_{Au}\sim (4\ts 10^{-3})^{-1/2}V_{As}\sim 16 V_L$, so from  $(11)$
$$V_{rec}=V_{u} \sim r_t/\tau_{rec}\sim r_t/(\tau_{ed} F(R_M)/100)\sim
V_L,\eqno(12)$$
where the last equality follows for Petschek reconnection with
$F(R_M)\sim {\rm Log}[R_M]\sim 25$
and using $(10)$.  The tube reconnection, which
occurs between regions with $\be_{t}<<1$, can in principle
proceed with $V_{rec}\sim V_L$ even for $\beta_{ave}\sim 1$.
That $V_{rec}\sim V_L$ supports the use of
$(11)$ even if tubes fill a large volume fraction.

Note that flux tube mobility requires that the turbulent damping viscosity be
negligible (Vishniac 1995).
This condition is  $\nu_T < V_L r_t/\pi^3$
where $\nu_T$ is the viscosity.
For $\beta_{ave}\sim 1$, this requires
$R_L\equiv V_L L/\nu_T>\pi^3 (L/r_t)\sim 500$.
$R_L\sim 1000$ was used above.

$\bf IV.\ Application\ to\ Extragalactic\ Sources$




Solving the $in\ situ$ acceleration problem by the flux tube
scheme requires that
1) $(8)$ is satisfied and that 2) electrons encounter a
reconnection acceleration site within their synchrotron loss times.
Consider requirement 1:
Since $V_{u} \sim V_{rec}\sim V_L\sim V_b$, using $(12)$ and EQP, and
$B_u\sim B_{t}$, we can use $(8)$ and  $(2)$ applied to flux tubes to
obtain the new condition
$$1.6\ts 10^{4}(\nu/GHz)^{1/2} \lsim 10^8(V_b/10^8{\rm\ cm/s}){\rm
Cos} \theta, \eqno(13)$$
which is satisfied for reasonable choices of
jet and lobe parameters.

Requirement 2 means that the distance between sites
must be less than $D_{syn}$, or more conservatively,
that every line of sight must pass through
a  reconnection region.  Since 1/2 of any tube of length $L$ is
incurring reconnection for $\beta_{ave}\sim 1$ as described above,
we have $D_J>2L\sim 8r_t$
where $D_J$ is the jet diameter.  This
provides an upper limit on the size of the turbulent outer
scale given by $L<D_J/2$.  If $D_J\sim 0.5{\rm kpc}$
for example, this implies $L\lsim 250$pc.  From $(6)$, this in turn implies
that $r_t\sim
15.6$pc.  The size scale of tubes in the jets
is a lower limit to the size scale of tubes in the lobes.

Because $L$ is much smaller than jet lengths of interest, a jet field
would have many reversals along its length.
This is consistent with observations in jets if EQP is
assumed (Begelman et al., 1984).
Field lines parallel to a jet can carry a magnetic flux
$\lsim 10^{34}$G cm$^2$ near a central black hole of $10^9M_\odot$
and then $\lsim 10^{37}$G cm$^2$ at 1pc.  However, the observed EQP
field strengths and cross sectional areas at $\gsim$ 10 kpc
imply more like $10^{40}$G cm$^2$.  Thus the field likely reverses many
times, consistent with the flux tube scheme.


The total luminosity produced by the flux tube model is given by
$$L_{tot}=N_{rr}L_{rr},\eqno(14)$$
where $L_{rr}$ is the luminosity per reconnecting region.
Near EQP, $L_{rr}\sim 4 \times 10^{-3}
(B_u^2/8\pi)W_{rr}(V_{rec}/r_t)$.
Then using $(10)$ and $(12)$, $(14)$ becomes
$$L_{tot}\sim 4\times 10^{-3}
(B_t^2/8\pi)W_{tot}(10/\tau_{ed}),\eqno(15)$$
where $W_{tot}$ is the total volume of the region of interest.
For a lobe of diameter 1Mpc, $B_t\sim 5\mu$G,
and $V_b\sim 10^7$cm/s,
 $\tau_{ed}\sim L/V_L\sim 250{\rm pc}/ V_b\sim 7.5\times
10^{13}$sec so
$L_{tot}\sim 5 \ts 10^{45}$erg/sec.
For kpc scale jets with length 10 kpc, width 0.5 kpc,
$100$GeV electrons, $B_t\sim 10^{-3}$G, and  $V_b\sim V_L\sim 0.1c$,
$L_{tot}\sim 10^{44}$erg/sec.
These are within the range of observed luminosities for
powerful radio sources (Muxlow \& Garrington, 1991).

$\bf V.\ Conclusions$

That slow shocks from reconnecting flux tubes might be able to
provide the {\it in situ}
acceleration of electrons required in
large scale jets and radio lobes is suggested.
If a steady-state flux tube structure ensues, reconnection regions will occur
naturally throughout the turbulent plasmas, and
therefore, so will the associated slow shocks.  Even if
$\beta_{ave} \gsim 1$, the flux tubes
are low $\beta$ regions.  The slow shocks may then Fermi accelerate
electrons to power law spectra reasonably
consistent with the observed ranges for radio sources.  The standard
electron injection problem for  shocks may be overcome by direct
injection from the X-point regions.  Future work should
include a more dynamical approach, and a consideration of relativistic bulk
flows.


\noindent 1)  Eqn. $(9)$ differs from Vishniac (1995) by a factor of 2.

\noindent {\bf Acknowledgments}:
I would like to thank G. Field, E. Vishniac, \& I. Yi for
related discussions.

\noindent Achterberg, A., 1987,
in $Astrophysical\ Jets\ and\ Their\ Engines$, W. Kundt ed.,
NATO series C, vol 208, (Dodrecht:  Reidel).

\noindent  Begelman,M.C., Blandford,R.D., \& Rees,M.J., 1984,
Rev. Mod. Phys., $\bf 56$ 255.

\noindent Birkinshaw, M., 1991, in $Beams\ and\ Jets\ in\
Astrophysics,$ ed. P.A. Hughes,  (New York: Cambridge Univ. Press).

\noindent Biskamp,D., 1994, Phys.  Rep., $\bf 237$, 179.

\noindent Blackman,E.G. \& Field, G.B., 1994, Phys Rev. Lett, $\bf
73$, 3097.

\noindent Coroniti,F.V. et al., (1994),  J. Geophys. Res., $\bf 99$, 11251.


\noindent Eilek,J.A., \& Hughes, P.A., 1991, in $Beams\ and\ Jets\ in\
Astrophysics,$ ed. P.A. Hughes,  (New York: Cambridge Univ. Press).

\noindent Ellison,D.C., 1992, in $Particle\ Acceleration\ in\ Cosmic\
Plasmas$, G.P. Zank \& T.K. Gaisser eds. (New York:  American Institute of
Physics).

\noindent Feldman,W.C. et al., 1990, J. Geophys. Res., $\bf 90$, 233.


\noindent Heiles,C., 1995, in {\it Physics of the Interstellar Medium and
Intergalactic Medium}, A. Ferrara ed., (San Francisco: Astronomical
Society of the Pacific).

\noindent Hoshino, M. et al., 1992, Ap. J., $\bf 390$, 454.

\noindent Jones,F.C., \& Ellison,D.C., 1991, Space Sci. Rev., $\bf
58$, 259.

\noindent Kantrowitz, A., \& Petschek,H.E., 1966  in $Plasma\ Physics\
in \ Theory\ and\ Application$
edited by W. B. Kunkel, (New York: McGraw-Hill).

\noindent Landau,L.D. \&  Lifshitz,E.M., 1987, $Fluid\
Mechanics$, (Oxford:  Pergamonn Press).

\noindent Macchetto,F., in $Science\ With\ The\ Hubble\ Space\
Telescope$,
eds. P. Benvenuti \& E. Schreier, ESO Conference Proc. $\bf 44$ (1992).

\noindent Meisenheimer,K., \& R$\ddot{\rm o}$ser,H.J., 1986, Nature
$\bf 319$, 459.

\noindent Muxlow,T.W.B., \& Garrington,S.T., 1991, in $Beams\ and\ Jets\ in\
Astrophysics,$ ed. P.A. Hughes,  (New York: Cambridge Univ. Press).


\noindent Omidi, N. \& Winske,D., 1994,
J. Geophys. Res., $\bf 97$, 14801.

\noindent Parker, E.N., $Cosmical\ Magnetic\ Fields$, 1979,
(New York:  Oxford University Press).

\noindent Romanova,M.M., \& Lovelace,R.V.E., 1992, A.\& A., {\bf 262}, 26.

\noindent Rybicki, G.B. \& Lightman, A.P. 1979,
Radiative Processes in Astrophysics, (New York: John Wiley and Sons).

\noindent Vishniac,E.T., 1995, Ap. J., $\bf 446$, 724.
\end



\end